\documentclass[a4paper]{jpconf}
\usepackage{graphicx}

\newcommand{\eps} {\epsilon}

\newcommand{\tanb} {\tan\beta}

\newcommand{\hone} {h_{1}}
\newcommand{\htwo} {h_{2}}

\newcommand{\ra} {\rightarrow}
\newcommand{\nbtag}  {N_{b-tag}}
\def\met        {E\!\!\!\!/_T}

\newcommand{\bea}{\begin{eqnarray}}
\newcommand{\eea}{\end{eqnarray}}
\newcommand{\beq} {\begin{equation}}
\newcommand{\eeq} {\end{equation}}

\begin{document}
\title{CP-violating MSSM Higgs at Tevatron and LHC}

\author{Siba Prasad Das$^{1}$ and Manuel Drees$^{2}$}

\address{$^1$ 
AHEP Group, Institut de F\'{\i}sica Corpuscular --
  C.S.I.C./Universitat de Val{\`e}ncia \\
  Edificio Institutos de Paterna, Apt 22085, E--46071 Valencia, Spain,}
\address{$^2$ Bethe Center for Theoretical Physics and Physikalisches
  Institut, Universit\"at Bonn, Nussallee 12, D--53115
  Bonn,  Germany}
\ead{spdas@ific.uv.es, drees@th.physik.uni-bonn.de}

\begin{abstract}

  We analyze the prospect for observing the intermediate neutral Higgs boson
  ($h_2$) in its decay to two lighter Higgs bosons ($h_1$) at the presently
  operating hadron colliders in the framework of the CP violating MSSM using
  the PYTHIA event generator. We consider the lepton+ 4-jets+ $\met$ channel
  from associate $W h_2$ production, with $W h_2 \ra W h_1 h_1 \ra \ell
  \nu_\ell b \bar b b\bar b$. We require two or three tagged $b$-jets. We
  explicitly consider all relevant Standard Model backgrounds, treating
  $c$-jets separately from light flavor and gluon jets and allowing for
  mistagging. We find that it is very hard to observe this signature at the
  Tevatron, even with 20 fb$^{-1}$ of data, in the LEP--allowed region
  of parameter space due to the small signal efficiency. At the LHC, a priori
  huge SM backgrounds can be suppressed by applying judiciously chosen
  kinematical selections. After all cuts, we are left with a signal cross
  section of around 0.5 fb, and a signal to background ratio between 1.2 and
  2.9. According to our analysis this Higgs signal should be viable at the LHC
  in the vicinity of present LEP exclusion once 20 to 50 fb$^{-1}$ of data
  have been accumulated at $\sqrt{s}=14$ TeV.

\end{abstract}


\section{Introduction} 
\label{intro}
 
The Minimal Supersymmetric Standard Model (MSSM) \cite{MSSM} requires
two Higgs doublets, leading to a total of five physical Higgs bosons.
At the tree level, these can be classified as two neutral CP--even
bosons ($\phi_1$ and $\phi_2$), one neutral CP--odd boson ($a$) and
two charged bosons. In the presence of CP violation, the three neutral
Higgs bosons can mix radiatively \cite{CPmixing0,Lee:2003nt}. The mass
eigenstates $h_1$, $h_2$ and $h_3$ with $m_{h_1} < m_{h_2} < m_{h_3}$
can then be obtained from the interaction eigenstates $\phi_1$, $\phi_2$
and $a$ with the help of the orthogonal matrix $O_{\alpha i}$,
$(\phi_1,\phi_2,a)^T_\alpha= {O_{\alpha i}}(h_1,h_2,h_3)^T_i \, ,$
which diagonalizes the Higgs boson mass matrix. $O$ depends on various
parameters of the SUSY Lagrangian.

Due to this mixing, the Higgs mass eigenstates are no longer CP
eigenstates.  Moreover, the masses of the Higgs bosons, their
couplings to SM and MSSM particles, and their decays are significantly
modified \cite{Lee:2003nt}. For example, the Higgs boson couplings to
pairs of gauge bosons are scaled by $g_{h_iVV}$ relative to the
SM. These couplings can be expressed as $g_{h_iVV} = \cos\beta\,
O_{\phi_1 i}\: +\: \sin\beta\, O_{\phi_2 i} \, ,$ where $\tan\beta$ is
the ratio of Higgs vacuum expectation values (VEVs). The magnitude of
$g_{h_2 W W}$ is directly related to the production process studied in
this paper.

In the absence of mixing between neutral CP--even and CP--odd states
the LEP experiments were able to derive absolute lower bounds of about
90 GeV on the masses of both the lighter CP--even Higgs and the
CP--odd boson \cite{adlo}.  However, in the presence of CP violation,
the LEP experiment were not able to exclude certain scenarios with
very light $h_1$. In this ``LEP hole'' $h_1$ is dominantly a CP--odd
state with almost vanishing coupling to the $Z$ boson. One then has to
search for $Z h_2$ or $h_1 h_2$ production. In part of the LEP hole,
these cross sections are suppressed by the rather large $h_2$
mass. Moreover, $h_2 \ra h_1 h_1$ decays lead to quite complicated
final states, which often yield low efficiencies after cuts.  One
LEP--allowed region has $m_{h_1}$ less than 10 GeV, so that $\hone \ra
\tau^+ \tau^-$ is dominant; in the other, $m_{h_1} \sim 30 - 50$ GeV
so that $\hone \ra b \bar b$ is dominant. $m_{h_2}$ lies between
slightly below 90 and slightly above 130 GeV. Scenarios with even
lighter $h_2$ are excluded by decay--independent searches for $Z h_2$
production \cite{adlo}. If $m_{h_2}$
is much above 130 GeV, the CP--odd component of $h_1$ becomes
subdominant, so that the cross section for $Z h_1$ production becomes
too large. Finally, the LEP hole occurs for $\tan\beta$ in between 3
and 10 \cite{adlo}. We analyze the prospect for 
observing a signal for the production of neutral Higgs bosons in the 
second of these LEP allowed regions. 

\section{Numerical analysis}
\label{numerical}

In our analysis we took five different benchmark points, denoted by S1 through
S5 [$m_{h_2}$=130-90 GeV with $m_{h_1}$=30 GeV] and two $\rm
CPX_{0.5}$--scenario [$\rm CPX$--1(2) where $m_{h_2}$=102 (103) GeV with
  $m_{h_1}$=36 (45) GeV, which can be realized for $M_{H^+}$ = 131.8 GeV with
  $\tanb$ = 4.02 (4.39)] of the MSSM~\cite{Carena:2000ks}.  We calculated the
spectrum and the couplings for these two benchmark points using {\tt
  CPsuperH}~\cite{Lee:2003nt}.  In our simulation we used the {\tt PYTHIA
  v6.408 } \cite{pythia} event generator with the {\tt SLHA}
\cite{Skands:2003cj} interfacing option.  We used {\tt MadGraph/MadEvent
  v4.2.8} \cite{Maltoni:2002qb} for generating parton level SM backgrounds
which were fed to {\tt PYTHIA} for showering. We set the renormalization and
factorization scale to $Q= \sqrt {\hat s}$ and used CTEQ5L for the parton
distribution functions (PDF).

The signal arises from $p \bar p \rightarrow W \htwo \rightarrow \ell
\nu_{\ell} \hone\hone \rightarrow \ell \nu_{\ell} b \bar b b \bar b$, leading
to $\ell jjjj \met$ events, where $\ell = e$ or $\mu$. The effective
cross section for this signal topology can be expressed as,
\beq  \label{effcross}
\sigma_{\rm signal}^{\rm tot} = \sigma_{SM}(p \bar p/pp \rightarrow W
\htwo) \times   g^2_{h_2 W W}  \times Br(\htwo \ra \hone\hone) \times
Br(\hone \ra b \bar b)^2  \times 2{Br(W \rightarrow e \nu_{e})}\ ,  
\eeq
where $g_{h_2WW}$ is the $h_2 WW$ coupling in units of the corresponding SM
value, $W$ stands for $W^\pm$ and the factor 2 is for $\rm \ell = e$ and
$\mu$. This process has recently been studied in refs.\cite{Koreans,Han},
using parton--level analyses with quite promising results.  We instead
performed a full hadron--level analysis, including initial and final state
showering as well as the underlying event. We take $g^2_{h_2WW} \times Br(h_2
\rightarrow h_1 h_1) \times Br(h_1 \rightarrow b \bar b)^2 = 0.50$.

We simulate our signal and backgrounds at Tevatron Run-II with $\sqrt s =1.96$
TeV. We have used the toy calorimeter simulation ({\tt PYCELL}) provided in
{\tt PYTHIA} with the following criteria: calorimeter coverage is $\rm |\eta|
< 3.64$; the segmentation is given by $\Delta \eta \times \Delta \phi$=$0.16
\times 0.098$ which resembles the CDF detector \cite{Aaltonen:2007dm};
Gaussian smearing of the total energy of jets and leptons; a cone algorithm
with $\rm\Delta R(j,j) = \sqrt{\Delta\eta^{2}+\Delta\phi^{2}} = 0.4 $ has been
used for jet finding; $\rm E_{T,min}^{cell} \ge 1.5$ GeV is considered to be a
potential candidate for jet initiator; minimum summed $\rm E_{T,min}^{jet} \ge
10.0$ GeV is accepted as a jet and the jets are ordered in $E_{T}$; leptons
($\rm \ell = e, ~\mu$) are selected with $\rm E_T^{\ell} \ge 15.0$ GeV and
$\rm |\eta^{\ell}| \le 2.0$ and no jet should match with a hard lepton in the
event. We reconstructed the missing energy ($\met$) from all observed
particles.

The tagging of $b-$jets plays a crucial role in our analysis. A jet with
$\left|\eta_j \right| \le 1.2$ and $E_{T}^j \geq 15$ GeV ``matched'' with a
$b-$flavored hadron ($B-$hadron), i.e.  with $\Delta R(j,B-{\rm hadron}) <
0.2$, is considered to be ``taggable''. We assume that such jets are actually
tagged with with probability $\eps_b = 0.50$ \cite{Hanagaki:2005fz}. We find
that our tagging algorithm agrees well with the $t \bar t$ analysis of CDF
\cite{cdfttbar}.  We also modeled mistagging of non$-b$ jets as $b-$jets,
treating $c-$jets differently from those due to gluons or light quarks with
$\eps_c = 0.10$ and $\eps_{u,d,s,g} = 0.01$.

\begin{figure}[h]
\begin{minipage}{18pc}
\includegraphics[width=18pc]{tev_h2_3b.epsi}
\caption{\label{tev_mh2_3b}  
The four--jet invariant mass $m_{4j}$ distribution after all
cuts for signal scenarios S1, S3 and S5 and for the total background
(ToB), requiring triple $b-$tag following 
the last column of Table~\ref{tab:eventTeVLHC} at Tevatron.}
\end{minipage}\hspace{2pc}%
\begin{minipage}{18pc}
\includegraphics[width=18pc]{lhc_h2_3b.epsi}
\caption{\label{lhc_mh2_3b} 
The four--jet invariant mass $m_{4j}$ distribution after all
cuts for signal scenarios S1, S3 and S5 and for the total background
(ToB), requiring triple $b-$tag following 
the last column of Table~\ref{tab:eventTeVLHC} at LHC.}
\end{minipage}
\end{figure}

The cross-sections for the signal benchmarks lie in between $0.045-0.16$ pb
while the total background is approximately $455$ pb.  We have displayed the
raw number of events for signal and backgrounds in the first column in Table
~\ref{tab:eventTeVLHC}. The second column is the number of events after
applying the following basic acceptance cuts: $N_{\rm jet} \ge 4$,
$E_{T}^{j=1-4} > 10$ {\rm GeV}, $|\eta^{j=1-4}|$$<$$3.0$; $N_{\rm lepton} \ge
1$,$E_{T}^{\ell}$ $>$ 15 {\rm GeV}, $|\eta^{\ell}|$$<$$2.0$ and $\met \geq$ 15
{\rm GeV}. Not surprisingly, $t \bar t$ is the main source of background at
this stage.  We found that the suppressions by applying $\nbtag \geq 2$ are
almost the same for signal and $t \bar t$. The signal contains more $b$
quarks, but the $t \bar t$ background has much harder $b$ jets, leading to
larger tagging probabilities. However, this background can contain a third $b$
jet only due to showering. Hence requiring $\nbtag \geq 3$ greatly reduces the
background and increases the signal--to--background ratio. Unfortunately the
small triple $b-$tagging probability, which lies between 1.3 and 3.6\%
depending on $m_{h_2}$, also reduces the signal rate. 

We have also looked for invariant mass peaks to isolate the signal on top of a
sizable background.  In addition to the basic acceptance cuts (defined above)
we demand that the signal contains exactly (rather than at least) four
jets. This reduces combinatorial backgrounds for Higgs mass
reconstructions. Finally, we pick the jet pairing $(ij) (kl)$ (with $i,j,k,l
\in \{1,2,3,4\}$) that minimizes the difference $m_{j_ij_j} - m_{j_kj_l}$ of
di--jet invariant masses; in the absence of showering and for perfect energy
resolution, the signal would have $m_{j_ij_j} = m_{j_kj_l} = m_{h_1}$. We then
demand that both $m_{j_ij_j}$ and $m_{j_kj_l}$ lie between 10 and 60 GeV,
where the lower bound comes from the requirement that $h_1 \rightarrow b \bar
b$ decays should be allowed, and the upper bound from the requirement that
$h_2 \rightarrow h_1 h_1$ decays should be open. We next require the four--jet
invariant mass to lie between 60 and 140 GeV; this covers the entire ``LEP
hole'' in the MSSM Higgs parameter space. Table~\ref{tab:eventTeVLHC} shows
that after these cuts, the signal actually exceeds the total background.

\begin{table}
\caption{\label{tab:eventTeVLHC} Process column shows the signal benchmarks
  and the SM backgrounds, where $j$ stands for $u,d,s,g$ and -- for background
  process 7 and 8 -- $c$.  RawEvt is stands for the number of events produced
  in the experiments (for backgrounds we applied basic pre--selections in the
  generator level: $p_{T}^{j,b} \geq 5$ {\rm GeV}, $|\eta^{j,b}|$$<$$5.0$ and
  $\Delta R (jj,bb,bj) \geq 0.3$). $N_{acc}$ is the number after the basic
  selection cuts, whereas $N_{3b}$ is with at least three jets tagged as
  $b-$jets, allowing for mistagging. Eff3 is the number of events passing the
  selection cuts that contain exactly four jets and at least 3$b-$ tagged
  jets; the numbers in parentheses represent the number of events with the
  inclusion of $m_{\rm pair}$ and $m_{4j}$ cuts. Finally, ToB is the total
  number of background events.}

\begin{center}
\begin{tabular}{|l|llll||llll|}
\hline
 & \multicolumn{4}{c||}{Tevatron with 4fb $^{-1}$ } & 
\multicolumn{4}{c|}{LHC with 10fb $^{-1}$} \\
\br
Process& RawEvt & $N_{acc}$ & $N_{3b}$&Eff3 (+h1) &RawEvt&$N_{acc}$&$N_{3b}$&Eff3(+h1)\\
\mr
S1 & 38.11 & 11.09 & 0.77 &0.49 (0.40)& 1157 &352.5&33.48&13.96(6.85)\\
S2 & 51.59 & 13.85 & 0.83 &0.51 (0.44)& 1486 &418.3&36.84&15.36(7.76)\\
S3 & 68.91 & 16.58 & 0.83 &0.54 (0.47)& 1962 &506.5&39.81&17.03(8.91)\\
S4 & 94.76 & 19.88 & 0.87 &0.54 (0.46)& 2620 &610.6&43.18&17.81(9.61)\\
S5 & 133.6 & 23.92 & 0.86 &0.52 (0.45)& 3516 &724.7&43.92&18.96(9.63)\\
CPX-1&89.89& 20.27 & 0.82 &0.49 (0.43)& 2509 &600.2&40.07&17.25(9.07)\\
CPX-2&87.56& 22.46 & 0.84 &0.53 (0.47)& 2421 &597.2&40.28&16.88(9.64)\\
\hline
$t \bar t$ &6760&3545&25.62&9.65(0.05)&1,690,000&818,800&7795&1469(5.52)\\
$b \bar b b \bar b W^{\pm}$&12.59&1.52&0.06&0.03(0.01)&337.6&31.8&4.10 &2.95(0.53)\\ 
$b \bar b b j W^{\pm}$ &0.043 &0.01&0&0(0)&23.3&2.3&0.13&0.11(0.01)\\ 
$b \bar b c j W^{-}$ &131.2&17.6&0.05&0.02(0.01)&73,170&7359&77.56&56.50(6.79)\\ 
$b \bar b c \bar c W^{\pm}$&44.51&5.53&0.03 &0.01(0.01)&1126&89.9&1.68&1.17(0.25)\\ 
$b \bar b j j W^{\pm}$ &5181&610.2&0.28&0.15(0.06)&535,700&45,830&17.14&17.89(1.93)\\ 
$b jj j W^{\pm}$ &14.31&2.01&0&0(0)&7194&586.3&0.23&0.05(0.01)\\ 
$jjjj W^{\pm}$ &384000&47340&0.24&0.02(0)&59,700,000&4,332,000&2.18&4.59(0.68) \\ 
$t \bar t b \bar b $ &12.15&6.80&0.44 &0.03(0)&10,100&5700&751.5&72.82(1.28)\\  
$t \bar t c \bar c $ &21.69&13.66&0.19 &0.02(0)&16,440&9245&259.8&31.54(0.45)\\
\hline
ToB &396200 &51550 &26.91 &9.93(0.14) &62,030,000 & 5,220,000 & 8910 &1657(17.45)\\  
\br
\end{tabular}
\end{center}
\end{table}

The results of the column labeled Eff3 in Table~\ref{tab:eventTeVLHC} 
have been obtained by including all events that pass the other cuts 
(not related to tagging) and have at least three {\em taggable} jets, 
assigning each event a weight given by its (mis)tagging probability. 
This greatly increases the statistics. 
We checked that this gives results that are consistent with the event
rejection technique whenever the latter has good statistics; 
this is the case if at most one $b-$tag results from mistagging.

The distribution of $m_{4j}$ is shown in Fig.~\ref{tev_mh2_3b} for signal
scenarios S1, S3 and S5 as well as for the total background. We observe clear
peaks for the signal, which are shifted downwards due to, for example,
showering (by 10 to 15 GeV) from the naive expectation $m_{4j} = m_{h_2}$ at
the parton level.

By seeing the distributions of $m_{\rm pair}$ (average of the two optimal
pairing of di--jet invariant masses) and $m_{4j}$ (see Fig.~\ref{tev_mh2_3b})
allow us define the final significance of the signal by counting events that
satisfy: $0.6 m_{h_1} \leq m_{\rm pair} \leq m_{h_1} + 5 \ {\rm GeV}$ and $0.7
m_{h_2} \leq m_{4j} \leq m_{h_2} + 10 \ {\rm GeV}$.  We see that requiring
triple $b-$tags leads to very good signal to background ratio, of around 10
for $m_{h_1} = 30$ GeV and slightly less for heavier $h_1$. However, we expect
less than 2 signal events after all cuts for a total integrated luminosity
($\int {\cal L} dt$) of 20 fb$^{-1}$. We tried also to see the effect for two
$b-$tag and found that the signal rate increase by a factor between 3.7 and 5
while the background increases by two orders of magnitude and $S/\sqrt{B}$ is
well below two.

The significance defined in this way overestimates the true statistical 
significance of a double peak in the $m_{\rm pair}$ and $m_{4j}$ distributions
somewhat, due to the ``look elsewhere'' effect: since $m_{h_1}$ and
$m_{h_2}$ are not known a priori, one would need to try different
combinations when looking for peaks. However, given that we use rather
broad search windows, there are probably only ${\cal O}(10)$
statistically independent combinations within the limits of the LEP
hole. 
Note also that the signal rate is still quite small. Further
kinematical cuts, which might slightly increase the signal to
background ratio, are therefore not likely to increase the statistical
significance of the signal. We are therefore forced to conclude that
the search for $W h_2 \rightarrow W h_1 h_1 \rightarrow \ell \nu b
\bar b b \bar b$ events at the Tevatron does not seem promising, and
turn instead to the LHC.

Our analysis for the LHC follows broadly similar lines as that for the
Tevatron. We simulate our signal and backgrounds at the LHC with $\sqrt s =
14$ TeV. The {\tt PYCELL} model is based on the ATLAS detector \cite{atlas}
with calorimeter coverage $\rm |\eta| < 5.0$, segmentation $\Delta \eta \times
\Delta \phi = 0.087 \times 0.10$. We use a Gaussian energy resolution for
leptons and jets and a cone algorithm for jet finding, with $\rm\Delta R(j,j)
= 0.4$.  Calorimeter cells with $\rm E_{T,min}^{cell} \ge 1.0$ GeV are
considered to be a potential candidates for jet initiator. All cells with $\rm
E_{T,min}^{cell} \ge 0.1$ GeV are treated as part of the would--be jet;
minimum summed $\rm E_{T,min}^{jet} \ge 15.0$ GeV is accepted as a jet and the
jets are ordered in $E_{T}$; leptons ($\rm \ell = e, ~\mu$) are selected if
they satisfy $\rm E_T^{\ell} \ge 20$ GeV and $\rm |\eta^{\ell}| \le 2.5$.  The
jet--lepton isolation criterion and the missing transverse energy $\met$ are
adopted similar to our Tevatron analysis.

Only jets with $|\eta_j| < 2.5$ are considered to be taggable as
$b-$jets. If the jet is ``matched'' to a $b-$flavored hadron, with
$\Delta R(j,{\rm hadron}) \leq 0.2$, the tagging efficiency is taken
to be 50\%. If instead the jet is matched to a $c-$hadron, the
(mis)tagging efficiency is taken to be 10\%, whereas jets matched to
a $\tau-$lepton have zero tagging probability. All other taggable jets
have (mis)tagging probability of 0.25\%. These efficiencies follow
recent ATLAS analyses~\cite{Aad:2009wy,Lehmacher:2008hs,mistaggcharm}.

The cross-sections for the signal benchmarks lie between $0.5-1.7$ pb while
the total backgrounds is approximately $2.84 \cdot 10^4$ pb. We have displayed
the raw number of events for signal and backgrounds (again with the same
generator--level cuts as of Tevatron) in the first column in Table
~\ref{tab:eventTeVLHC}. The second column is the number of events after
applying the basic acceptance cuts: $N_{\rm jet} \ge 4$, $E_{T}^{j=1-4} > 15$
{\rm GeV}, $|\eta^{j=1-4}|$ $<$$5.0$; $N_{\rm lepton} \ge 1$, $E_{T}^{\ell}$
$>$ 20 {\rm GeV } $|\eta^{\ell}|$$<$$2.5$ and $\met \geq$ 20 {\rm GeV}. They
reduce the cross section by about a factor of 5 (3) for $m_{h_2} = 90 \ (130)$
GeV.

The number of events (for $\int{\cal L} dt$=10 fb$^{-1}$) passing the
acceptance cuts and containing exactly four jets, at least three of which are
tagged (adopting the same strategy like Tevatron), is given by the Eff3 column
of Table~\ref{tab:eventTeVLHC}.  Similar to Tevatron we require both jet pair
invariant masses to lie between 10 and 60 GeV and $m_{4j}$ to lie between 60
and 140 GeV. The $m_{4j}$ distribution is shown in Fig.~\ref{lhc_mh2_3b}.
After these cuts we are left with slightly less than one signal event and
slightly less than two background events per fb$^{-1}$ of data. A $5 \sigma$
signal would then require almost 100 fb$^{-1}$ of data, more than the LHC is
likely to collect during ``low'' luminosity running.

We also checked that requiring a fourth $b-$tag reduces the signal 
cross section by another order of magnitude or more. The signal rate 
then becomes so low that one would have to wait for the high--luminosity 
phase of the LHC to accumulate enough events to reconstruct invariant 
mass peaks. We therefore stick to triple $b-$tag in our LHC analysis.

We found that the background shows a peak in the $m_{\rm pair}$ distribution
between 30 and 40 GeV, not far from the peak of the signal in the scenarios we
consider. A tighter cut on $m_{\rm pair}$ will nevertheless improve the
signal--to--background ratio. Moreover, the four--jet invariant mass
distribution (in Fig.~\ref{lhc_mh2_3b}) of the background peaks at large
values, largely due to the contribution from $t \bar t$ production.  At least
for scenarios with $h_2$ masses in the lower half of the ``LEP hole'' region a
tighter cut on $m_{4j}$ will therefore also improve the significance of the
signal. We therefore applied the same ``double--peak'' cuts as at the
Tevatron. The signal then always exceeds the background.  Assuming an
integrated luminosity of 60 fb$^{-1}$ at the end of ``low'' luminosity
running, we find a final statistical significance of at least 5 standard
deviations, and a signal sample of some 30 events.

\section{Conclusions} 

We analyzed the possibility of observing neutral Higgs bosons at currently
operating hadron colliders in the framework of the CP violating MSSM. We
explored the $\ell jjjj \met$ channel with double, triple and quadruple $b$
tag, focusing on the region of parameter space not excluded by LEP
searches. We considered a large number of SM backgrounds and employed a full
hadron--level Monte Carlo simulation using the {\tt PYTHIA} event generator.
We carefully implemented $b-$tagging, including mistagging of $c-$jets or
light flavor or gluon jets.  At the Tevatron, requiring 3$b-$tag, we can only
expect about one signal event per 10 fb$^{-1}$ of $\int{\cal L} dt$, on a
background of about 0.3 events. If we require only 2$b-$tag, the signal
increases by a factor of about 4, but the background increases by two orders
of magnitude, making the signal unobservable.  At the LHC we focussed on
events with exactly four jets, cutting simultaneously on the average di--jet
invariant mass and the four--jet invariant mass and demanding at least
3$b$-tags. We found a signal rate above the background, and a signal
significance exceeding 5$\sigma$ for an $\int{\cal L} dt$ of 60 fb$^{-1}$.
One might be able to increase the S:B ratio even more by requiring 4$b-$tags
with softer tagging criteria (enhancing mistag rate also), possibly
simultaneously relaxing the requirement on the number of jets. This could be
used to confirm the existence of a signal.

We conclude that searches for $W h_2$ production at the LHC with $W
\rightarrow \ell \nu$ and $h_2 \rightarrow h_1 h_1 \rightarrow b \bar b b \bar
b$ should be able to close that part of the ``LEP hole'' in parameter space
where $h_1 \rightarrow b \bar b$ decays dominate. The details of this analysis
can be found in our recent paper \cite{dd}.

\subsection{Acknowledgments}
This work was partially supported by the Bundesministerium f\"ur Bildung 
und Forschung (BMBF) under Contract No. 05HT6PDA, by the EC contract 
UNILHC PITN-GA-2009-237920, and by the Spanish grants FPA2008-00319, 
CSD2009-00064 (MICINN) and PROMETEO/2009/091 (Generalitat Valenciana).

\smallskip
\end{document}